 \documentclass[aps,prb,twocolumn,groupedaddress,showpacs,preprintnumbers,amsmath,amssymb]{revtex4}
\usepackage{graphicx}
\usepackage{dcolumn}
\usepackage{bm}

\begin{document}

\title{Short-Range Order in Fe-Rich Fe-Cr Alloys as Revealed by M\"ossbauer   Spectroscopy}

\author{S. M. Dubiel} \email[Corresponding author: ]{dubiel@novell.ftj.agh.edu.pl} \author{J. Cieslak} \affiliation{Faculty of   Physics and Applied Computer Science, AGH University of Science and   Technology, al. Mickiewicza 30, 30-059 Krakow, Poland}

\date{\today}

\begin{abstract}
  The distribution of Cr atoms in Fe$_{100-x}$Cr$_x$ alloys with $x \le 25$   within the first two coordination shells, $1NN-2NN$, around probe $^{57}$Fe
  atoms was studied using M\"ossbauer Spectroscopy. Clear evidence was found   that the distribution is not random but
instead characteristic of a given   atomic configuration, $(m,n)$ ($m$ being the number of Cr atoms in $1NN$,   and $n$ that in $2NN$).
The behavior was described quantitatively in terms   of average SRO parameters, $<\alpha_1>$ (for $1NN$), $<\alpha_2>$
(for   $2NN$) and $<\alpha_{12}>$ (for $1NN-2NN$) as well as in terms of a local   short-range order (SRO) parameter, $\alpha(m,n)$,
for each pair $(m,n)$. A   change of sign (inversion) was found both in $<\alpha_1>$ and in   $<\alpha_2>$, though going with $x$ in opposite directions.
 No inversion was   observed in $<\alpha_{12}>$, which was either positive or negative   depending on the metallurgical state of the samples.
These findings   prompt a revision of current interpretation of experimental and theoretical   results relevant to the issue.
\end{abstract}


\maketitle

Among various binary alloys of iron, Fe-Cr alloys occupy a special role for both scientific and technological reasons. They can be treated as good systems
for testing models and theories, especially those relevant to magnetism where different phases are exhibited depending on alloy composition. \cite{Burke83}.
 Their crystallographic structure, which for many years was regarded as homogenous {\it bcc} over the whole concentration range, turned out to be much more
 complex when a tetragonal $\sigma$-phase and a miscibility gap were discovered. The latter two phenomena are on one hand of interest {\it per se} and have
been the subject of intensive study \cite{Xiong10}, and on the other hand, are also of a great importance technologically, namely in the production of
important grades of stainless steels \cite{Lo09} for which the Fe-Cr alloys are the basic ingredient. Consequently, their useful properties such as a good
 resistance to high-temperature corrosion and good mechanical properties (toughness, ductility and welding ability) may be severely degraded if the
$\sigma$-phase precipitates or phase separation into Fe-rich and Cr-rich phases occurs.

Recently there has been increased interest in Fe-Cr alloys.  This is driven both by the discovery of giant magnetoresistance in Fe/Cr layers \cite{Gruenberg86} and also by the potential for Fe-Cr-based steels to be used in the construction of a new generation of power plants -- advance fusion and fission reactors or high power accelerator spallation targets \cite{Mansu04}. In the latter application, the materials undergo irradiation damage which can seriously degrade their mechanical properties. On the lattice scale, the radiation causes lattice defects, and, consequently, a redistribution of Fe/Cr atoms that can result in a short- range order (SRO) or phase decomposition into Fe-rich and Cr-rich phases.

According to previous neutron diffraction (ND) studies
\cite{Mirebeau84,Mirebeau10}, the Cowley SRO-parameter, $<\alpha_{12}>$, was
found to change its sign at $x\approx 10-11$.  This finding was qualitatively
confirmed by a M\"ossbauer spectroscopic (MS) study \cite{Filippova00}, yet
the value of the critical concentration was not determined. Additionally,
theoretical calculations predicted the existence of such an inversion, but for
different values of $x$ \cite{Caro05,Lavrentiev07}.  The aim of the present
investigation was to study the issue in more detail using MS, since this method
applied to the Fe-Cr alloys can provide precise and relevant information on
SRO for each statistically meaningful atomic configuration, $(m,n)$, where $m$
is the number of Cr atoms in the first-nearest neighbor shell, $1NN$, and $n$
is the number in the second-nearest neighbor shell, $2NN$
\cite{Dubiel74,Dubiel81}. Such information would be much more
detailed than the one recently found with ND \cite{Mirebeau10}, where the
inversion of the SRO parameter was found as the average over the $1NN-2NN$
volume.

There are 63 different atomic configurations possible for the {\it bcc}
structure within such volume. Although for a random distribution the
probability of most of them, $P(m,n)$, is very small, all those with
$P(m,n)>\sim0.01$ are measurable using MS. This improvement over the
information available by ND means that MS can be used as a more adequate basis
for quantitative verification of different theoretical models pertinent to the
issue \cite{Caro05,Lavrentiev07,Turchi94,Wallenius04}.

$^{57}$Fe M\"ossbauer spectra were recorded at 295 K in transmission mode on
four series of Fe$_{100-x}$Cr$_x$ alloys, I, II, IIIa and IIIb, with
different histories and composition using a standard spectrometer with a
sinusoidal drive and a $^{57}$Co/Rh source of 14.4 keV gamma rays.

Samples of series I, with $0 \le x \le 15$, were 40 years old. They were
prepared as follows: Armco iron and 4N - purity chromium were melted in a
vacuum induction furnace. After melting, they were kept in a liquid state for
about 10 minutes and cooled down to an ambient temperature.  The ingots were
next forged into flat bars (8 mm thick), which were subsequently cold-rolled
into 2 mm thick tapes.  The tapes were annealed in a vacuum at 840$^\circ$C for
1h, and then cooled in a furnace.  The 2 mm thick tapes were next
rolled down to a thickness of 0.1 mm from which 20-30 $\mu$m thick foils were
obtained again by cold rolling. Samples of series II, three years old and with
$15 \ge x \le 25$, were prepared in a similar way.  Examples of the spectra
recorded on these samples are shown in Fig. 1.

The series IIIa and IIIb samples were EFDA/EURATOM model Fe-Cr alloys that had
been prepared in 2007.  They were delivered in the form of bars 10.9 mm in
diameter, in a recrystallized state after cold reduction of 70\% and then heat
treated for 1h under pure Ar flow at the following temperatures: 750$^\circ$C for
Fe$_{94.4}$Cr$_{5.6}$, 800$^\circ$C for Fe$_{89.75}$Cr$_{10.25}$ and
850$^\circ$C for Fe$_{85}$Cr$_{15}$ followed by air cooling.  For the MS
measurements, a slice $\sim$1 mm thick was cut off from each bar using a
diamond saw, and was subsequently cold-rolled down to a final thickness of
20-30 $\mu$m.  Samples obtained in this way constituted the series IIIa. Some of the latter in the form of 20mm-diameter circular foils
were annealed at 800 $^\circ$C for 4h under Ar flow followed by a liquid nitrogen
quenching. The samples that underwent this heat treatment constituted the
series III. All the spectra were analyzed in the same way i.e. with
the two-shell model. It was assumed that only Cr atoms situated within the
$1NN$ and $2NN$ neighbor-shells cause measurable changes in spectral
parameters i.e.  the hyperfine field, $B$, and the isomer shift, $IS$.  It was
also assumed that the changes both in $B$ and in $IS$ were additive
i.e. $X(m,n;x) = X(0,0;x)- m \Delta X_1-n \Delta X_2$, where $X = B$ or $IS$
and $\Delta X_{12}$ stands for the change in $X$ due to one Cr atom in $1NN$
(index 1) or in $2NN$ (index 2).

This procedure proved to be successful in the analysis of the spectra of various Fe-rich Fe-X alloys (e.g. Ref.15).
 Each spectrum was treated as composed of a number of subspectra, $N$, corresponding to a particular atomic configuration, $(m,n)$.
Its relative spectral area was equal to the probability of the atomic configuration associated with the spectrum, $P(m,n)$.
The latter is the relevant quantity for a quantitative description of a real atoms distribution over lattice sites in a given sample.
Theoretically, the probabilities for the random case, $P_r (m,n;x$), can be calculated from the formula:

\begin{figure}[tb]
  \includegraphics[width=.35\textwidth]{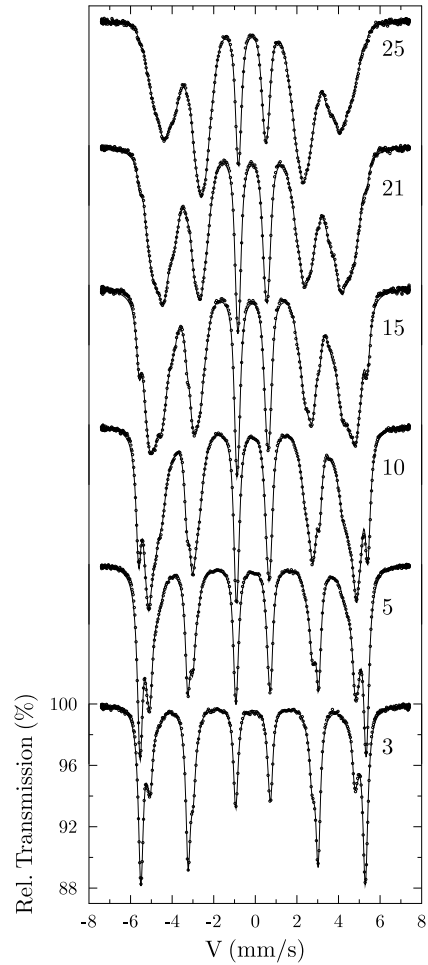}
  \caption{Examples of $^{57}$Fe M\"ossbauer spectra recorded at 295 K on
    Fe-Cr alloys with various Cr content in at\% (3=3.25, 5=4.85, 10=10.25,
    15=14.9, 21=21.0, 25=25.0). The solid lines are equations of best fit.  }
  \label{fig1}
\end{figure}

\begin{equation}
  P_r(m,n;x)=\left(
    \begin{array}{c}
      8 \\ m
    \end{array}
  \right)
  \left(
    \begin{array}{c}
      6 \\ n
    \end{array}
  \right)
  x^{m+n}(1-x)^{14-m-n}
  \label{eq1}
\end{equation}

\begin{figure}[tb]
  \includegraphics[width=.45\textwidth]{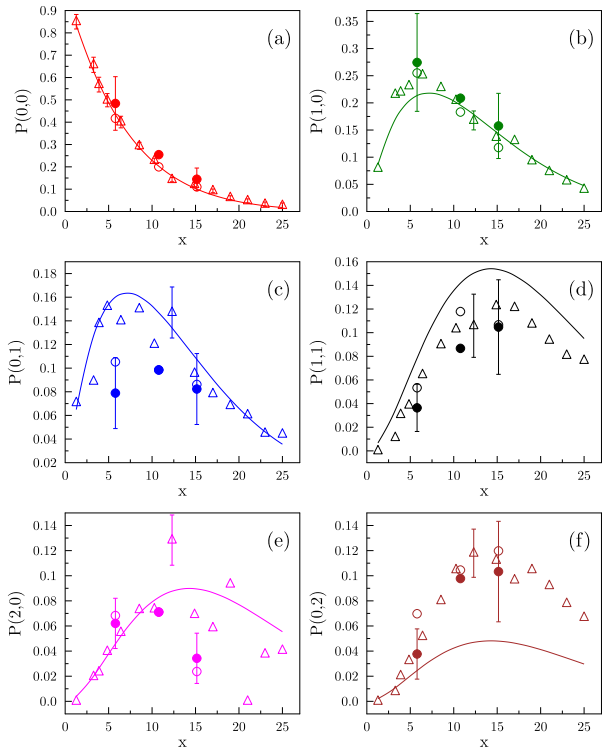}
  \caption{(Color online) Probabilities of various atomic configurations,     $P(m,n)$, around the probe $^{57}|$Fe atoms as calculated for a random
   distribution (solid lines) and as derived from the spectra (open triangles    series I and II, open circles series IIIa, full circles series IIIb).  }
  \label{fig2}
\end{figure}

Most of the possible 63 configurations have vanishingly small probabilities,
and can therefore be neglected. In practice, one usually takes into account
only the most probable ones in order to fulfill the condition: $\sum
P(m,n;x)>0.99$.  This condition significantly reduces $N$ from 63 to e. g. $N
=4$ ($x=1$), $N = 10$ ($x = 10$), $N = 14$ ($x =15$).  Using the above
described procedure, we have successfully fitted all the measured spectra with
the following values of the spectral parameters: $\Delta B_1 = 31.0\pm0.5$
kOe, $\Delta B_2 = 21.3\pm0.6$ kOe, $\Delta IS_1 = -0.022\pm0.001$ mm/s and
$\Delta IS_2 = -0.009\pm0.001$ mm/s which agree well with those previously
reported \cite{Dubiel74,Dubiel81}. The values of $P(m,n)$ determined from the
analysis for the most significant 6 configurations are presented in Fig. 2
together with the corresponding ones calculated from formula (1). It is clear
that the actual distribution is, in general, not random, and the degree and
direction of deviation from randomness is characteristic of a given atomic
configuration. In particular, the $P(0,0)$ values are close to the $P_r(m,n)$
ones for all $x$ values, though those determined for the series IIIb show a
systematic deviation. The distribution of atoms is partly random for $(0,1)$,
$(1,0)$ and $(2,0)$ configurations, and non-random for $(1,1)$ and $(0,2)$
configurations.  In order to quantitatively describe the actual departure from
the randomness for $(m,n)$, we introduce the following measure for the
short-range order:

\begin{equation}
  \alpha(m,n) = \frac{P(m,n)}{P_r(m,n)}-1
  \label{eq2}
\end{equation}

This SRO parameter can be regarded as an adequate measure for the departure of
the actual distribution of atoms from the random distribution, and its value
can be easily determined based on the spectral parameters. $\alpha(m,n) > 0$
when the actual probability of finding an $(m,n)$ atomic configuration around
the probe Fe atom is higher than the one for the random distribution (atomic
short-range ordering), and $\alpha(m,n) < 0$ otherwise (clustering). In our
opinion this definition is simple and justified, at least from the viewpoint
of MS, as it ascribes a positive value of $\alpha$ to a larger number of Cr
atoms within $1NN-2NN$ shell, as seen by the probe Fe atoms, than the one
expected for the random case, and a negative value otherwise.

\begin{figure}[bt]
  \includegraphics[width=.45\textwidth]{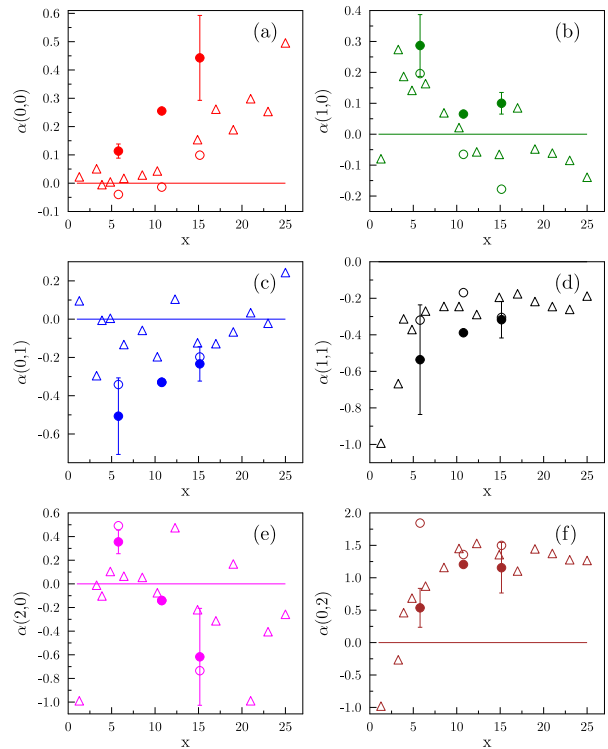}
  \caption{ (Color online) SRO parameters, $\alpha(m,n)$, for various atomic     configurations versus Cr content, $x$,
as calculated using formula (\ref{eq2}).  Open triangles stand for series I and II, open circles for series IIIa, and full circles for series IIIb.  }
  \label{fig3}
\end{figure}

Values of $\alpha(m,n)$ obtained from formula (2) for all 4 series are
presented in Fig. 3.  Here it is clearly evident that the $\alpha(m,n)$'s are
characteristic of a given $(m,n)$ and that they also depend on the samples'
histories. It is evident then that the actual distribution of atoms in the Fe-Cr
system is much more complex than the one obtained from the ND experiments
\cite{Mirebeau84,Mirebeau10}. The inversion of the SRO parameter at $x\approx
10$, can locally (i.e. in terms of $(m,n)$) only be seen for the following
configurations: (1,0) series I and IIIa, (2,0) series IIIa and IIIb, and
perhaps (0,0) series IIIa. The opposite inversion takes place at $x\approx
3$ in $\alpha(0,2)$.  On the other hand, $\alpha(1,1) < 0$ over the whole
concentration range showing a saturation behavior. It is also worth noticing
that $\alpha(0,0) \approx 0$ for $x<\sim 10$, but only for the cold-rolled
samples.

An important issue is the effect of heat treatment on the distribution. To get
some insight one can compare the $\alpha(m,n)$'s calculated for series IIIa and IIIb.  Interestingly, for some configurations viz. (0,0) and (1,0)
they are significantly different, while for other viz. (0,1), (2,0) and (1,1) they are similar.  In these circumstances, it seems reasonable to also
introduce average values of $\alpha$ to the description of the actual distribution of Cr atoms in the studied samples.  Thus, the average $\alpha$ for
the $1NN$ shell, $<\alpha_1>$, that for the $2NN$ shell, $<\alpha_2>$, and also the average for the $1NN-2NN$ shells, $<\alpha_{12}>$, can be defined
 as follows:

\begin{equation}
  \label{eq3}
  <\alpha_i> = \frac{<k>}{<k_r>}-1
\end{equation}

where $k=m,n,m+n$ for $i=1, 2, 12$, respectively, and $<m>$ is the average number of Cr atoms in $1NN$, $<n>$ is that in $2NN$, and $<m+n>$ is that
in $1NN-2NN$ as determined from analysis of the M\"ossbauer spectra. The three symbols with subscript $r$ represent the same quantities but calculated
 for the random distribution. A graphical illustration of $<\alpha_{1}>$, $<\alpha_{2}>$ and $<\alpha_{12}>$ is displayed in Fig. 4 on the left-hand
side of the panel.  Alternatively, following Wittle and Campbell \cite{Wittle85}, and staying with the $1NN-2NN$ model and {\it bcc} structure, one
can define $<\alpha_{1}>$, $<\alpha_{2}>$ and $<\alpha_{12}>$ as follows:

\begin{equation}
  \label{eq4}
  <\alpha_i> = \frac{<k>-lx}{l(1-x)}-1
\end{equation}
where $k=m,n,m+n$ and $l=8,6,14$ for $i=1, 2, 12$, respectively.  It should be remembered that both definitions of the average $\alpha$ differ
in sign in comparison to that introduced by Cowley.

\begin{figure}[bt]
  \includegraphics[width=.45\textwidth]{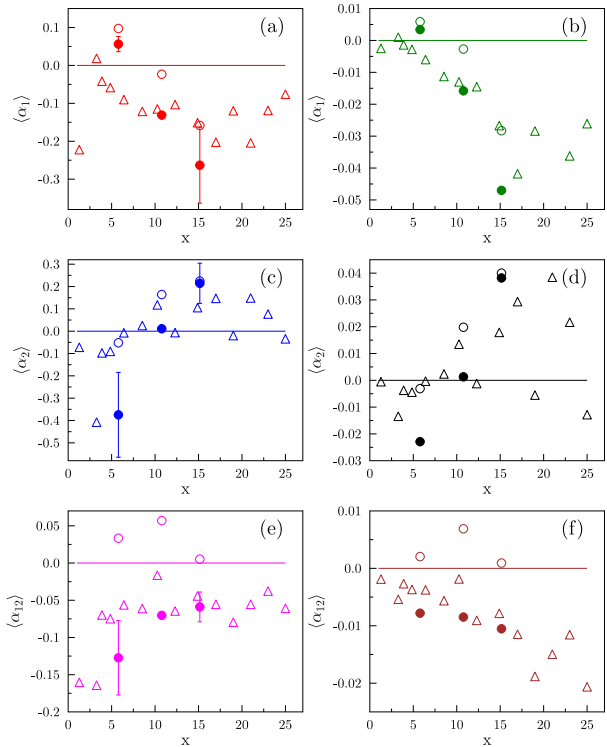}
  \caption{ (Color online) The average SRO parameters for the $1NN$ shell,     $<\alpha_1>$, for the $2NN$ shell, $<\alpha_2>$, and that for the
   $1NN-2NN$ shells, $<\alpha_{12}>$, versus Cr concentration, $x$, as     calculated using formula (\ref{eq3}) -- left-hand panel -- and using
   formula (\ref{eq4}) -- right-hand panel.  Open triangles stand for the     series I and II, open circles for the series IIIa, and full circles for the
  series IIIb.  }
  \label{fig4}
\end{figure}

The average SRO parameters obtained using formula (\ref{eq4}) are presented on the right-hand side panel of Fig. 4. Quantitative agreement between the
corresponding average SRO parameters obtained with the two approaches can be readily seen.  Concerning the crucial question of the inversion, one can
definitely observe its existence both in $<\alpha_{1}>$ and $<\alpha_{2}>$, especially in the samples of series IIIa and IIIb.  However, the inversions
 go in opposite directions: on increasing $x$ one observes a change from ordering to clustering in the former and a change from clustering to ordering
in the latter. The critical concentration at which the inversion occurs depends on the samples' histories. As a consequence of such behavior, the SRO
parameter averaged over the $1NN-2NN$ volume, $<\alpha_{12}>$, does not show any inversion: the one for series IIIa is positive, hence revealing the
short-range ordering with a maximum at $x \approx 10$, while that for series IIIb is negative, hence indicating the clustering effect. Similar effect
was revealed for a series of Au-Fe, where the alloys were found to exhibit either clustering of Fe atoms or atomic short-range ordering depending on
 their metallurgical state and heat treatment \cite{Wittle84}. The presently reported behavior is completely different than the one found with
ND \cite{Mirebeau84,Mirebeau10}.

To summarize, the distribution of Cr atoms in the $1NN-2NN$ shells was studied
quantitatively using M\"ossbauer spectroscopy on the level of atomic
configurations in 4 series of Fe-Cr with having different metallurgical
states.  Clear evidence was found that the actual distribution is much more
complex than the one accepted to date from ND \cite{Mirebeau84,Mirebeau10}. In
particular, it was shown here that the change of the SRO-parameter sign observed with ND at 10-11 at\% Cr and regarded as an experimental confirmation
 of various theoretical calculations including SRO itself \cite{Lavrentiev07}, is exclusively a feature of the $1NN$ shell.  The average SRO parameter
for the $2NN$ shell was also found to exhibit inversion but in the opposite direction, i.e. from clustering to ordering.  Consequently, the SRO parameter
 averaged over the two shells, $<\alpha_{12}>$, does not show any inversion and its actual value depends on the metallurgical state of the samples: for
the cold-rolled ones its is positive, hence indicative of the atomic short-range ordering, whereas for the quenched samples it is negative, indicating
the existence of clustering.
In other words, the actual distribution of atoms in the Fe-Cr alloys is very sensitive to their metallurgical state. This, in turn,
reflects the fact that the initial state of these alloys is metastable. Upon heating the alloys decompose into Fe-rich and Cr-rich phases.
The degree of the decomposition, hence the actual distribution of atoms (and values of the SRO-parameters) depends on samples' metallurgical histories as experimentally revealed in this
study, and theoretically demonstrated by performing Monte Carlo atomistic simulations \cite{Erhart08}.
\begin{acknowledgments}
  This work, supported by the European Communities under the contract of   Association between EURATOM and IPPLM, was carried out within the framework
 of the European Fusion Development Agreement, and it was also supported by   the Ministry of Science and Higher Education, Warszawa. J. Aguilar is
thanked for his help in improving the language.
\end{acknowledgments}


\begin{thebibliography}{58}



\bibitem{Burke83} \bibinfo{author}{S.K. Burke}, \bibinfo{author}{R. Cywinski},   \bibinfo{author}{J.R. Davis} and \bibinfo{author}{B.D. Rainford},   \bibinfo{journal}{J. Phys. F: Metal Phys.}, \textbf{\bibinfo{volume}{13}},   \bibinfo{pages}{451} (\bibinfo{year}{1983}).

\bibitem{Xiong10} \bibinfo{author}{W. Xiong}, \bibinfo{author}{M. Selleby},   \bibinfo{author}{Q. Chen}, \bibinfo{author}{J. Odqvist} and   \bibinfo{author}{Y. Du}, \bibinfo{journal}{Rev. Solid State Mater. Sci.}   \textbf{\bibinfo{volume}{35}}, \bibinfo{pages}{125} (\bibinfo{year}{2010}).

\bibitem{Lo09} \bibinfo{author}{K. H. Lo}, \bibinfo{author}{C. H. Shek} and   \bibinfo{author}{J. K. L. Lai}, \bibinfo{journal}{Materials Sci. Eng. R:     Reports} \textbf{\bibinfo{volume}{65}}, \bibinfo{pages}{39}   (\bibinfo{year}{2009}).

\bibitem{Gruenberg86} \bibinfo{author}{P. Gruenberg},   \bibinfo{author}{R. Schreiber}, \bibinfo{author}{Y. Pang},   \bibinfo{author}{M. B. Brodsky} and \bibinfo{author}{H. Sowers},   \bibinfo{journal}{Phys. Rev. Lett.}  \textbf{\bibinfo{volume}{57}},   \bibinfo{pages}{2442} (\bibinfo{year}{1986}).


\bibitem{Mansu04} \bibinfo{author}{L. K. Mansu},   \bibinfo{author}{F. Rowcliffe}, \bibinfo{author}{R.K. Nanstad},   \bibinfo{author}{S.J. Zinkle}, \bibinfo{author}{W.R. Corwin} and   \bibinfo{author}{R.E. Stolleret}, \bibinfo{journal}{J.Nucl. Mater}   \textbf{\bibinfo{volume}{329-333}}, \bibinfo{pages}{166}   (\bibinfo{year}{2004}).

\bibitem{Mirebeau84} \bibinfo{author}{I. Mirebeau},   \bibinfo{author}{M. Hennion} and \bibinfo{author}{G. Parette},   \bibinfo{journal}{Phys. Rev. Lett.}  \textbf{\bibinfo{volume}{53}},   \bibinfo{pages}{687} (\bibinfo{year}{1984}).

\bibitem{Mirebeau10} \bibinfo{author}{I. Mirebeau} and   \bibinfo{author}{G. Parette}, \bibinfo{journal}{Phys. Rev. B}   \textbf{\bibinfo{volume}{82}}, \bibinfo{pages}{104203}   (\bibinfo{year}{2010}).

\bibitem{Filippova00} \bibinfo{author}{N. P. Filippova},   \bibinfo{author}{V. A. Shabashov} and \bibinfo{author}{A. L. Nikolaev},   \bibinfo{journal}{Phys. Met. Metallogr.}  \textbf{\bibinfo{volume}{90}},   \bibinfo{pages}{145} (\bibinfo{year}{2000}).

\bibitem{Caro05} \bibinfo{author}{A. Caro}, \bibinfo{author}{D. A. Crowson}   and \bibinfo{author}{M. Caro}, \bibinfo{journal}{Phys. Rev. Lett.}   \textbf{\bibinfo{volume}{95}}, \bibinfo{pages}{075702}   (\bibinfo{year}{2005}).


\bibitem{Lavrentiev07} \bibinfo{author}{M. Yu. Lavrentiev},   \bibinfo{author}{D. Drautz}, \bibinfo{author}{D. Nguyen-Manh},   \bibinfo{author}{T. P. C. Klaver} and \bibinfo{author}{S. L. Dudarev},   \bibinfo{journal}{Phys. Rev. B} \textbf{\bibinfo{volume}{75}},   \bibinfo{pages}{014208} (\bibinfo{year}{2007}).

\bibitem{Dubiel74} \bibinfo{author}{S. M. Dubiel} and   \bibinfo{author}{K. Krop}, \bibinfo{journal}{J. Phys. (Paris)}   \textbf{\bibinfo{volume}{35}}, \bibinfo{pages}{C6-459}   (\bibinfo{year}{1974}).

\bibitem{Dubiel81} \bibinfo{author}{S. M. Dubiel} and   \bibinfo{author}{J. Zukrowski}, \bibinfo{journal}{J. Magn. Magn. Mater.}   \textbf{\bibinfo{volume}{23}}, \bibinfo{pages}{214} (\bibinfo{year}{1981}).

\bibitem{Turchi94} \bibinfo{author}{P. E. A. Turchi},   \bibinfo{author}{L. Reinhard} and \bibinfo{author}{G. M. Stocks},   \bibinfo{journal}{Phys. Rev. B} \textbf{\bibinfo{volume}{50}},   \bibinfo{pages}{15542} (\bibinfo{year}{1994}).

\bibitem{Wallenius04} \bibinfo{author}{J. Wallenius},   \bibinfo{author}{P. Olsson}, \bibinfo{author}{C. Lagerstedt},   \bibinfo{author}{N. Sandberg}, \bibinfo{author}{R. Chakova} and   \bibinfo{author}{V. Pontikis}, \bibinfo{journal}{Phys.Rev. B}   \textbf{\bibinfo{volume}{69}}, \bibinfo{pages}{094103}   (\bibinfo{year}{2004}).


\bibitem{Dubiel84} \bibinfo{author}{S. M. Dubiel} and   \bibinfo{author}{W. Zinn}, \bibinfo{journal}{J. Magn. Magn. Mater.}   \textbf{\bibinfo{volume}{45}}, \bibinfo{pages}{298} (\bibinfo{year}{1984}).

\bibitem{Wittle85} \bibinfo{author}{G. L. Wittle} and   \bibinfo{author}{S. J. Campbell}, \bibinfo{journal}{J. Phys. F: Metal Phys.}   \textbf{\bibinfo{volume}{15}}, \bibinfo{pages}{693} (\bibinfo{year}{1985}).

\bibitem{Wittle84} \bibinfo{author}{G. L. Wittle} and   \bibinfo{author}{S. J. Campbell}, \bibinfo{journal}{Phys. Rev. B}   \textbf{\bibinfo{volume}{30}}, \bibinfo{pages}{5364} (\bibinfo{year}{1984}).

\bibitem{Erhart08}
  \bibinfo{author}{P. Erhart},
  \bibinfo{author}{A. Caro}
  \bibinfo{author}{M. Serrano de Caro}
  and
  \bibinfo{author}{B. Sadigh},
  \bibinfo{journal}{Phys. Rev. B}
  \textbf{\bibinfo{volume}{77}},
  \bibinfo{pages}{134206}
 (\bibinfo{year}{2008}).


\end{thebibliography}
\end{document}